\documentclass[journal]{IEEEtran}
\usepackage{amsmath,amsfonts}
\usepackage{algorithmic}
\usepackage{algorithm}
\usepackage{array}
\usepackage[caption=false,font=normalsize,labelfont=sf,textfont=sf]{subfig}
\usepackage{textcomp}
\usepackage{stfloats}
\usepackage{url}
\usepackage{xcolor}

\usepackage{soul,xcolor}

\usepackage{verbatim}
\usepackage[colorlinks=true, linkcolor=blue, urlcolor=blue, citecolor=blue]{hyperref}
\usepackage{graphicx}
\usepackage{cite}
\usepackage{amsmath, amssymb, amsfonts}

\begin{document}

\title{Semantic Communications in 6G: Coexistence, Multiple Access, and Satellite Networks}

\author{
        \IEEEauthorblockN{Ishtiaque Ahmed, Yingzhuo Sun, Jingwen Fu, Alper Köse, Leila Musavian, Ming Xiao and Berna Özbek}
        \IEEEauthorblockA{
        }
    \thanks{This work was supported by the European Union’s Horizon Europe MSCA-DN programme through the SCION Project under Grant Agreement No. 101072375 and by UK Research and Innovation under the UK government’s Horizon Europe funding guarantee [Grants No. EP/X027201/1].} \thanks{I. Ahmed and L. Musavian are with the School of Computer Science and Electronic Engineering, University of Essex, Wivenhoe Park, Colchester CO4 3SQ, United Kingdom, E-mails: \{ishtiaque.ahmed, leila.musavian\}@essex.ac.uk.} \thanks{Y. Sun and J. Fu and M. Xiao are with  the Department of Information Science and Engineering, Royal Institute of Technology (KTH), 10044 Stockholm, Sweden, E-mails: \{yingzhuo, jingwenf, mingx\}@kth.se.} \thanks{A. Köse and B. Özbek are with  the Department of Electrical and Electronics Engineering, İzmir Institute of Technology, 35430 İzmir, Türkiye, E-mails: \{alperkose, bernaozbek\}@iyte.edu.tr.} \thanks{This work has been submitted to the IEEE for possible publication. Copyright may be transferred without notice, after which this version may no longer be accessible.}
 
}




\maketitle

\begin{abstract}
The exponential growth of wireless users and bandwidth constraints necessitates innovative communication paradigms for next-generation networks. Semantic Communication (SemCom) emerges as a promising solution by transmitting extracted meaning rather than raw bits, enhancing spectral efficiency and enabling intelligent resource allocation. This paper explores the integration of SemCom with conventional Bit-based Communication (BitCom) in heterogeneous networks, highlighting key challenges and opportunities. We analyze multiple access techniques, including Non-Orthogonal Multiple Access (NOMA), to support coexisting SemCom and BitCom users. Furthermore, we examine multi-modal SemCom frameworks for handling diverse data types and discuss their applications in satellite networks, where semantic techniques mitigate bandwidth limitations and harsh channel conditions. Finally, we identify future directions for deploying semantic-aware systems in 6G and beyond.
\end{abstract}


\section{Introduction}
\IEEEPARstart{T}{he} advancements in wireless communication networks has resulted in an exponential growth of wireless subscribers which poses a challenge to the limited bandwidth. With post-pandemic situations, the escalation in the demands of wireless services is further expected to boost, requiring researchers to work on the development of efficient systems for 6G which enables intelligent resource allocation within the limited spectrum.

To overcome the limitations associated with conventional communication systems, particularly that of approaching the Shannon’s limit, Semantic Communication (SemCom) paradigm has been examined. Based on the transmission of semantic content rather than raw bits, SemCom enhances the Spectral Efficiency (SE) by mitigating the transmission overhead \cite{letaief2019roadmap}. This paradigm, however, is best suited than the conventional Bit-based Communication (BitCom) at lower Signal-to-Noise Ratio (SNR) requirements, as BitCom can recover the data with nearly exact precision at higher SNR regimes. It is therefore viable to deploy both BitCom and SemCom for next-generation communication networks. 

In \cite{guo2024survey}, SemCom has been classified into two types based on network topology, namely paired SemCom, and network SemCom. For attaining improved reliability, ultra-high transmission efficiency, and compatibility, paired SemCom is deployed which involves direct interaction between any two participating agents. The two sharing agents mutually learn through the Knowledge Bases (KBs) and achieve an "understand-before-transmit" capability, extracting and sending only the core semantic content through semantic-level reasoning. Networked SemCom, on the other side involves more than two agents constituting a network with shared objectives and where each agent collaboratively shares to and learns from multiple agents. This flexible approach enables several intelligent modern era applications, like digital twins, virtual reality, autonomous vehicles, and the Metaverse.

Deep Learning (DL) techniques have started to gain considerable attention in the design of communication systems. For extracting the underlying features in text transmission, a DL based SemCom system named as DeepSC has been provided in \cite{xie2021deep}. The DeepSC is equipped with jointly trained transmitter and receiver, comprising of semantic encoder/decoder and channel encoder/decoder. More specifically, the neural networks for semantic features extraction is trained on the  Biderctional Encoder Representations from Transformers (BERT) model \cite{devlin2018bert} to quantify the resemblance between the transmitted and the received sentences. The semantic similarity between the original and reconstructed signals can be estimated with a function relying on the structure of neural networks in the DeepSC model, along with the prevalent channel conditions. However, there is an associated drawback owing to the inherent limitation of neural networks in replicating the exact signal for SemCom, even though how efficiently the model is trained using large KBs \cite{xie2021deep}. Therefore, BitCom holds its importance in scenarios where exact signal is required to be reconstructed. One efficient approach could be the one that deploys both SemCom and BitCom within the same communication network such that the drawbacks associated with each paradigm can be counterbalanced. Such heterogeneous users communication could become one of the key enabler for next-generation networks. However, to enable the coexistence of both types of users, novel strategies need to be developed, especially for allocating optimal resources to each type of user such that different access schemes can be flexibly supported within the same network \cite{ding2024next}.

The rate at which semantic information is transmitted with a prescribed accuracy is known as the semantic rate, which depends on the semantic similarity function $\epsilon(K, \gamma)$ with a value ranging from zero to one and a sigmoid shape pattern \cite{yan2022resource}. This function quantifies the resemblance between the original signal at the transmitter and the reconstructed signal at the receiver. For the value of $\epsilon(K, \gamma)$ closer to one, the signal will increasingly resemble the original bit-based signal rather than its reconstructed semantic counterpart \cite{xie2021deep}. As expressed in \cite{yan2022resource}, the semantic rate for any semantic user is given as
\begin{equation}
S = \frac{WI}{KL} \epsilon(K, \gamma),
\label{semanticrate}
\end{equation}
where $W$ is the channel bandwidth, $I$ is the amount of semantic information in any transmitted message in units of semantic (suts), $K$ represents the average number of semantic symbols transmitted for each word through DeepSC \cite{xie2021deep}, $\gamma$ represents the received SNR, and $L$ denotes the number of words. In practice, the closed-form expression of $\epsilon(K, \gamma)$ is not available, so the generalized logistic approximation is adopted via data regression on DeepSC outputs. Specifically, for each $K$ the DeepSC tool is run over a grid of SNR values to obtain empirical $\epsilon(K, \gamma)$ samples. Empirical results show that $\epsilon(K, \gamma)$ never decreases as $\gamma$ grows, and its slope first increases then decreases. \color{black} In \cite{mu2022heterogeneous},  the data-regression method has been deployed to tractably approximate the values of $\epsilon(K, \gamma)$ for different values of $K$ and $\gamma$ by following the criterion of minimum mean square error for fitting the values with a generalized logistic function (common form of sigmoid function) as 
\begin{equation}
\epsilon (K, \gamma) \approx \epsilon_K (\gamma) \overset{\triangle}{=} A_{K,1} + 
\frac{A_{K,2} - A_{K,1}}{1 + e^{-(C_{K,1} \gamma + C_{K,2})}},
\label{logisticapprox}
\end{equation}
where the lower (left) asymptote, upper (right) asymptote, growth rate, and the mid-point parameters of the logistic function are respectively denoted by $A_{K,1}$, $A_{K,2}$, $C_{K,1}$, and $C_{K,2}$ for different values of $K$. We choose this form because it accurately captures the behavior of $\epsilon(K, \gamma)$ at low and high SNRs by providing smooth monotonic growth between defined asymptotes.

In \cite{yan2022resource}, a relationship for the conventional bit rate $R_{\text{B}}$ and its equivalent semantic rate $R_{\text{SB}}$ has been established as:
\begin{equation}
R_{\text{SB}} = R_{\text{B}} \frac{I}{\mu L} \epsilon_C, \label{equivalentsemanticrate} 
\end{equation}
where $\mu$ is the average number of bits per word, and $\epsilon_C$ denotes the semantic similarity function associated with bit user's rate determined on the basis of Shannon’s capacity formula.

The key contributions of this work are:
\begin{itemize}
  \item We explore SemCom for satellite networks with diverse data types through adaptive KB alignment and semantic relay architectures to overcome bandwidth limits and harsh channels.
  \item We compare the performance of SemCom and BitCom in a two‐user single antenna uplink NOMA scheme where the near user employs BitCom with a constant power and the far user switches between BitCom and SemCom by reusing the same resource block via on–off power control.
  \item We survey multi‐modal SemCom frameworks that fuse text, images, audio and video into unified semantic representations via central fusion and task‐oriented Joint Source-Channel Coding (JSCC).
  \item We outline key challenges pertaining SemCom for enabling intelligent and energy-efficient 6G networks.
\end{itemize}

The remainder of this paper is organized as follows. Section II investigates SemCom's transformative role in satellite networks, featuring Artificial Intelligence (AI)-driven semantic relays and KB alignment. Section III analyzes multiple access techniques for heterogeneous SemCom/BitCom coexistence, including performance trade-offs in uplink scenarios. Section IV explores multi-modal semantic communication frameworks and their challenges in handling diverse data types. Finally, Section V concludes with future research directions for semantic-aware 6G systems.

\section{Semantic Communication for Satellite Communication}
This section examines how SemCom, combined with AI-driven techniques, is transforming satellite communications through adaptive KBs and semantic relay architectures. Satellite communications face significant hurdles due to harsh channel conditions, constrained bandwidth, limited frequency resources, and restricted computational and power capabilities which are exacerbated by increasingly diverse data types. SemCom provides a promising avenue for addressing these challenges by concentrating on the extraction of essential features in the transmitted data. This approach discards redundant information, thus lowering bandwidth demands and enhancing resilience to low Signal-to-Interference-and-Noise ratios (SINR). In addition, SemCom frameworks can define tailored performance metrics for various data formats and fuse multimodal information into compact, high-level representations. Critical enabling technologies include JSCC for integrated and efficient encoding, as well as generative AI techniques such as diffusion models that not only produce data ideally suited for semantic-driven transmission but also offer strong resistance to channel impairments. Furthermore, these diffusion-based models can adaptively equalize the channel more effectively than traditional neural networks. Finally, adaptive deep semantic strategies enable real-time adjustments to compression and transmission processes, optimizing resource usage even under stringent satellite link constraints. Taken together, these innovations point toward a more robust, bandwidth-efficient, and intelligent satellite communication ecosystem.


The work in \cite{huang2024prospects} delves into the challenges of satellite communication, particularly in the highly dynamic and resource-constrained environments of 6G satellite networks. The study highlights the potential of SemCom to mitigate these challenges by leveraging AI-driven semantic extraction, which enables efficient data transmission through intelligent feature selection. Instead of focusing on traditional performance metrics like bit error rate and channel capacity, the authors advocate for a paradigm shift towards semantic accuracy, where the effectiveness of communication is measured by how well the transmitted information aligns with the receiver's understanding. It outlines several enabling technologies, including JSCC for improved data compression, generative AI models that dynamically adapt to varying communication scenarios, and diffusion models that enhance transmission robustness in low SINR conditions. Furthermore, the concept of multi-modal information fusion, which allows satellite networks to integrate various data types such as text, images, and sensor readings into a unified semantic representation has been explored. It has been concluded that while SemCom presents significant advantages, further research is needed to optimize KBs and enhance adaptability in real-world deployments.

In \cite{hassan2024semantic}, the role of SemCom in optimizing data transmission for Earth observation satellites has been investigated. These satellites generate vast amounts of real-time data essential for applications such as disaster relief and environmental monitoring. However, the transmission of high-resolution imagery and other sensor data is often constrained by the limited bandwidth of satellite downlinks. A novel framework has been presented by employing SemCom techniques to prioritize and transmit only the most relevant information, reducing communication overhead without compromising data integrity. A key innovation of this study is the formulation of a latency minimization problem with Semantic Quality-of-Service (SC-QoS) constraints, ensuring that the received data retains its essential meaning. To address this optimization problem, a hybrid solution that combines a meta-heuristic Discrete Whale Optimization Algorithm (DWOA) with a one-to-one matching game has been introduced for efficient resource allocation. The study also incorporates joint semantic and channel encoding to enhance the downlink sum-rate, ultimately demonstrating that SemCom can significantly reduce transmission latency while preserving high-quality data delivery. The findings suggest that integrating SemCom into Earth observation satellites can dramatically improve bandwidth efficiency and system resilience in Non-Terrestrial Networks (NTNs).

In \cite{guo2024semantic}, SemCom-aware end-to-end routing in large-scale Low Earth Orbit (LEO) satellite networks has been introduced to explore the implications of SemCom for routing. Given the scarcity of available spectrum and the increasing demand for real-time multimedia services, it has been argued that traditional routing algorithms, which rely solely on Shannon's classic information theory, are inadequate. Instead, they proposed a SemCom-aware Routing (SCR) strategy that integrates AI-based semantic extraction with network-layer routing decisions. Unlike conventional routing protocols that optimize purely for link capacity and delay, SCR ensures that transmitted data is compatible with the receiver’s semantic KB, minimizing redundant transmission and improving overall network efficiency. The study models this problem using Integer Linear Programming (ILP) and introduces a temporal graph representation to capture the dynamic topology and KB distribution of satellite networks. Through extensive simulations on the Starlink constellation, it has been demonstrated that SCR outperforms traditional routing schemes by significantly reducing end-to-end delay and improving SE. This work highlights the importance of rethinking routing strategies in SemCom-enabled satellite networks and emphasizes the need for cross-layer optimization to fully exploit the benefits of AI-driven semantic extraction.

In \cite{chen2024semantic}, semantic-enhanced LEO satellite communication system with Orthogonal Time Frequency Space (OTFS) modulation has been investigated for improved downlink transmission. The OTFS modulation has gained attention due to its robustness against high Doppler shifts and multipath fading, which are common in satellite communications. The study proposes a novel semantic-aware OTFS framework that combines semantic feature extraction with optimized waveform design to maximize transmission efficiency. By utilizing DL-based semantic encoders, the system selectively extracts and transmits only the most relevant components of an image or video stream, reducing the overall data volume while maintaining high reconstruction quality. In addition to that, a joint optimization scheme that dynamically adapts transmission parameters has been introduced based on the semantic content and channel conditions, ensuring optimal performance under varying satellite link constraints. Experimental results show that this system achieves substantial gains in SE and robustness against channel impairments compared to traditional OTFS-based transmission. This study underscores the potential of combining SemCom with advanced modulation schemes to enhance the performance of next generation NTNs.

In \cite{jiang2024semantic}, semantic satellite communication based on generative Foundation Model (FM) explores the role of generative AI in enhancing SemCom for satellite networks. A framework that utilizes large-scale FMs has been presented to generate, compress, and transmit semantically rich information, reducing the dependency on raw data transmission. By employing pre-trained generative models, the system can synthesize high-fidelity reconstructed signals of transmitted data at the receiver, significantly reducing bandwidth requirements. A key innovation of this study is the adaptive semantic encoding mechanism, which dynamically selects the most relevant features based on the receiver’s KB and communication constraints. Moreover, a knowledge-driven reinforcement learning algorithm has been introduced to optimize transmission strategies in real-time, ensuring efficient adaptation to changing network conditions. Simulation results demonstrate that the generative FM-based approach achieves superior compression efficiency and resilience against channel noise compared to traditional SemCom frameworks. The study concludes that generative AI has the potential to revolutionize satellite communication by enabling highly efficient and intelligent semantic transmission systems.

\begin{figure}[t]
\centering
\includegraphics[trim={0cm 0cm 0cm 0cm},clip,width=1\columnwidth]{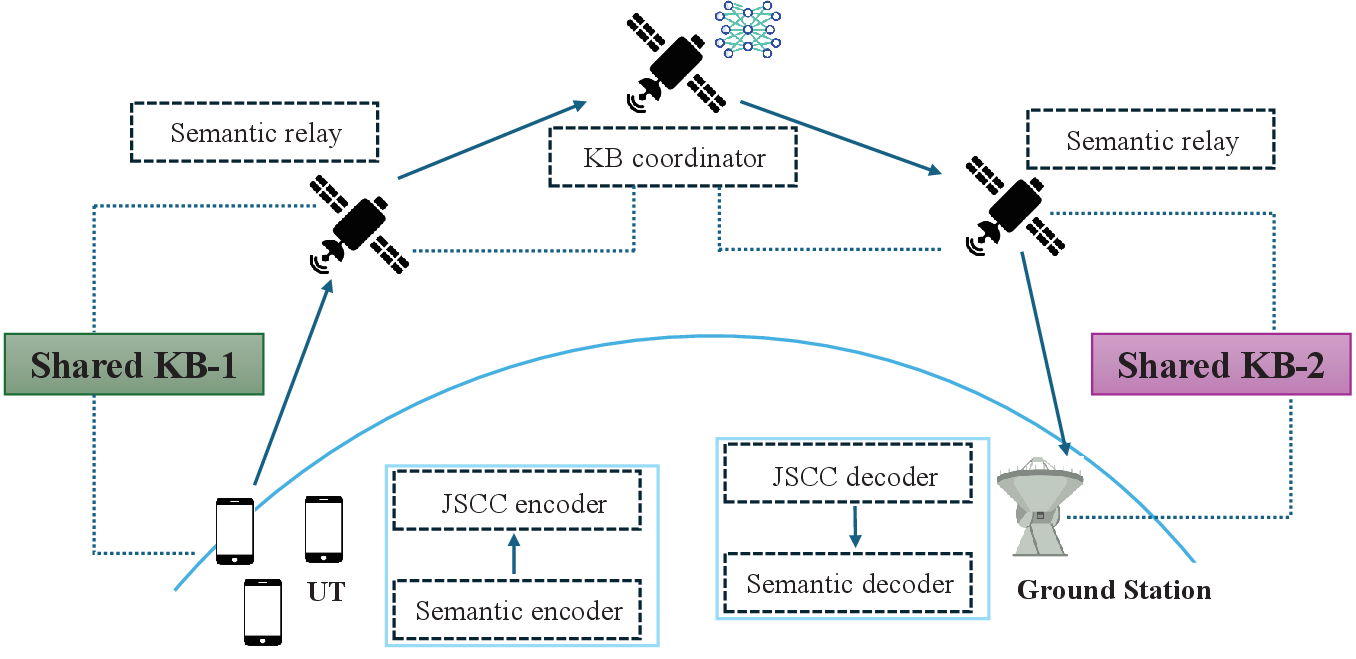}
\caption{Illustration of the semantic satellite relay communication system \cite{huang2024prospects}.}
\label{SemanticSat}
\end{figure}

Inspired by the existing literature, we propose a satellite communication system based on the semantic relay and semantic encoder/decoder. The shared KB design is vital for the SemCom system. However, it is difficult for the transmitter and the receiver to own the exactly same KB in the satellite communication scenario because of the wide coverage area and the large amount of nodes. Another challenge brought by the giant network is the severe cascaded fading. To overcome these two difficulties, the semantic relay satellite nodes and KB coordinator satellites \cite{huang2024prospects} can be used as illustrated in Fig.~\ref{SemanticSat}. The JSCC encoder and the semantic encoder are employed at the transmitter. The transmitted signal can be enhanced semantically based on the shared KB at the relay satellite. The enhanced signal will be transmitted to the KB coordinator satellite which can make the KBs at transmitter and receiver sides align with each other.

Collectively, a comprehensive perspective on the emerging field of SemCom in satellite networks illustrates how AI-driven semantic extraction, knowledge-based routing, advanced modulation techniques, and generative models can be leveraged to overcome traditional communication limitations. While significant progress has been made, several challenges remain, including the development of standardized semantic metrics, optimization of semantic KBs, and real-time adaptation to dynamic satellite network conditions. Focusing on refining these techniques to ensure seamless integration into next-generation satellite communication systems, ultimately will pave the way for more intelligent, efficient, and scalable global connectivity.

\section{Multiple Access Techniques for Next-Generation Networks} To enable massive connectivity, multiple access techniques play a critical role, especially to enable next-generation intelligent communication. One of the promising techniques enhancing the SE is the Non-Orthogonal Multiple Access (NOMA) where users share the same resource either through code or power-domain multiplexing. Although concerns relating to the fairness issues and increased complexity in implementation of Successive Interference Cancellation (SIC), pose challenges for NOMA, integrating it with SemCom can assist in the design of heterogeneous users communication networks \cite{mu2022heterogeneous}. We compare the performance of SemCom and BitCom for the two users uplink NOMA setup, with single antenna antenna each at the Access Point (AP) and users. Specifically, the user nearest to the AP communicates using traditional BitCom and is named primary user. The farther user termed as secondary, reuses the same resource block used by the primary user and can flexibly utilize either BitCom or SemCom based on the channel conditions for optimal communication policy. We assume the primary and secondary users are at the distance of 15 meters and 45 meters from the AP and are sharing the resource block via NOMA. 

For DeepSC with $K = 5$, the logistic approximation parameters are $A_{K,1} = 0.37$, $A_{K,2}= 0.98$, $C_{K,1}= 0.2525$, and $C_{K,2}= –0.7895$, providing high accuracy and a tractable form. To ensure high-quality semantic transmission, we set a threshold value of 0.9 for the semantic similarity-based reconstruction. The channel from each user to the AP is modelled as quasi-static block-fading, where it remains constant over a fading block and varies independently between blocks. Mathematically, the independently distributed Rayleigh fading channel coefficients are determined based on the distance-dependent path-loss model $\rho=\rho_0(1 / d)^\beta$. We assume a reference path-loss of $\rho_0=-30$ dB at 1 meter, and the path-loss exponent $\beta=4$, with $d$ denoting the link distance in meters. The AP experiences additive white Gaussian noise (AWGN) with zero-mean and -80 dBm variance. We want to maximize the ergodic semantic rate of the secondary user under the minimum ergodic rate constraint of the primary user. For BitCom, we set $\mu=40$ and $\epsilon_C = 1$ to represent error-free semantic reconstruction by the bit user. The ratio $I/L$ is set to a constant value of 1 in our simulations. \color{black} For a simpler analysis, we assume that the secondary user accesses the resources in an on-off approach by either transmitting with a constant power $P0$ or not accessing the resource at all. The primary user, however transmits with a constant power of 1 watt at all times. The results are generated using MATLAB and averaged over $10^5$ Monte Carlo channel realizations.

\begin{figure}[t]
\centering
\includegraphics[trim={0cm 0cm 0cm 0cm},clip,width=1\columnwidth]{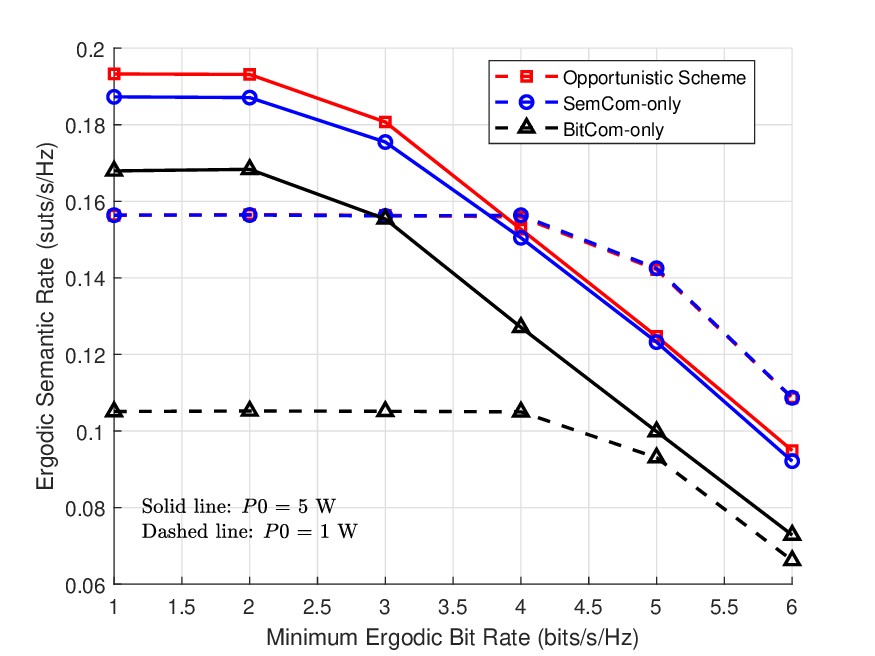}
\caption{Secondary user ergodic semantic rate versus minimum bit rate requirement at the primary user in two-user uplink NOMA.}
\label{fig1}
\end{figure}

Fig.~\ref{fig1} illustrates the ergodic semantic rate achieved per unit bandwidth by the secondary user versus the minimum ergodic bit rate of the primary user within the NOMA uplink setup. We compare the opportunistic (BitCom or SemCom) scheme at the secondary user with SemCom-only and BitCom-only schemes in a two-user scenario. Notably, as the required minimum ergodic bit rate of the primary user increases, the ergodic semantic rate of the secondary user decreases across all the compared schemes due to the limiting available resources for the secondary user with the stricter rate requirements at the primary user. Among the compared schemes, the opportunistic scheme consistently offers the highest ergodic semantic rates. This is due to the adaptive nature of the opportunistic scheme by which the best communication method is selected for higher spectral efficiency. Additionally, for transmit power $P0=1$ W at the secondary user, the SemCom-only scheme offers similar performance to the opportunistic scheme but significantly outperforms the BitCom-only scheme, reflecting the advantage of SemCom in low SNR regimes.

\begin{figure}[t]
\centering
\includegraphics[trim={0cm 0cm 0cm 0cm},clip,width=1\columnwidth]{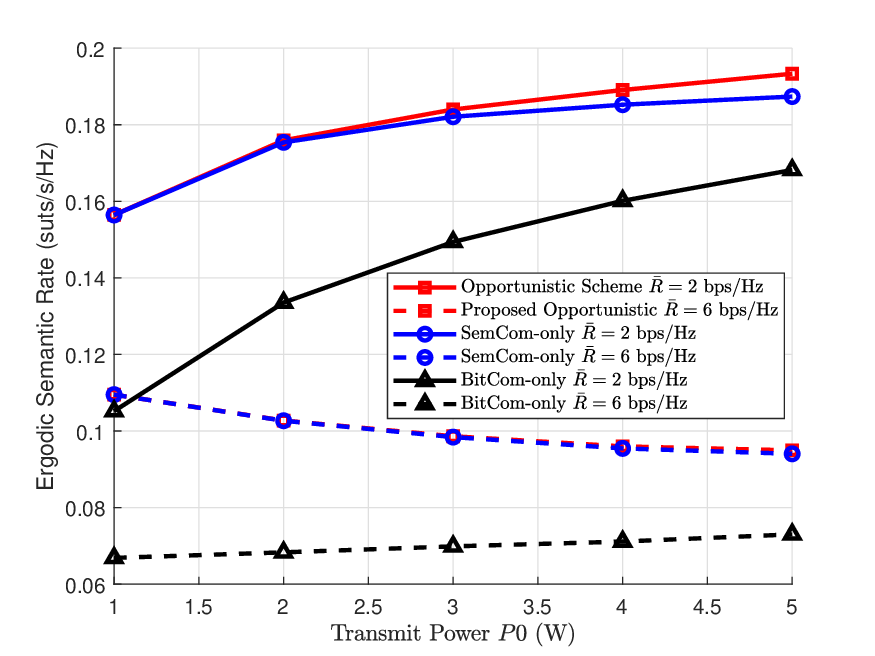}
\caption{Secondary user ergodic semantic rate versus $P0$ in two-user uplink NOMA.}
\label{fig2}
\end{figure}

Fig.~\ref{fig2} shows the ergodic semantic rate achieved at the secondary user versus $P0$ at the secondary user under minimum ergodic bit rate values ($\bar{R} = 2$ bps/Hz and $\bar{R} = 6$ bps/Hz) for the opportunistic, SemCom-only, and BitCom-only schemes. With increasing $P0$ values, the ergodic semantic rate across all schemes improves for the secondary user at $\bar{R} = 2$ bps/Hz. The opportunistic scheme offers best performance at both $\bar{R} = 2$ bps/Hz and $\bar{R} = 6$ bps/Hz minimum ergodic bit rate requirements, effectively selecting either SemCom or BitCom for communication. At $\bar{R} = 2$ bps/Hz, the performance of SemCom-only closely matches the opportunistic scheme, suggesting the effectiveness of SemCom in low-to-moderate bit-rate requirements. However, at $\bar{R} = 6$ bps/Hz, SemCom-only stops offering improvement with an increase in $P0$. The BitCom-only scheme shows significantly underperformance compared to the opportunistic and SemCom-only schemes, particularly at lower $P0$.


Another technique that can efficiently be deployed to integrate the SemCom and BitCom  is the hybrid NOMA which allows to capture the benefits of both Orthogonal Multiple Access (OMA) and traditional NOMA by allowing smooth transitions between these two schemes \cite{ding2024next}. Essentially, the hybrid NOMA enables multiple users to share the same resource block across different time slots, which ultimately enhances the overall system capacity. However, optimal power allocation and managing the complexities associated with SIC in hybrid NOMA framework remain important components for further investigation. Efficient power distribution can not only minimizes interference between bit-based and semantic users but also ensures that both user types achieve their required performance targets. By strategically allocating power among bit-based and semantic users, the system not only achieves an improved semantic SE with minimal interference between users but also ensures the minimum rate requirements for BitCom.

To address the issue of an efficient decoding mechanism when both types of users coexist, several methods are being investigated. If the semantic signal is to be decode first, it will be essential to ensure that the semantic decoding does not leave any residual interference that could affect the decoding of bit-based user. Therefore, bit-to-semantic decoding order will be a more beneficial choice in scenarios where both types of users coexist. In this case, the bit user will decode its signal first, with semantic signals subsequently decoded in an interference-free manner after SIC. Another effective strategy could be decoding the semantic signal using the shared KB between the semantic encoder and decoder, allowing the decoder to reconstruct missing or corrupted information by assessing semantic similarities. Additionally, AI-driven semantic decoding has gained attention due to the growing integration of DL techniques. Advanced models like transformers and autoencoders are employed to map transmitted messages to high-level semantic features.

While the integration of SemCom with conventional BitCom offers significant SE gains, the semantic paradigm extends far beyond single-modal data transmission. Recent advances have expanded SemCom’s scope to multi-modal frameworks—simultaneously processing text, images, audio, and video—enabling richer contextual understanding and task-oriented communication. This shift introduces new opportunities for efficiency but also demands innovative solutions to address challenges like cross-modal semantic fusion and bandwidth overhead, as explored in the following section.

\section{Multi-modal Semantic Communications}

\begin{figure}[t]
\centering
\includegraphics[trim={0cm 0cm 0cm 0cm},clip,width=1\columnwidth]{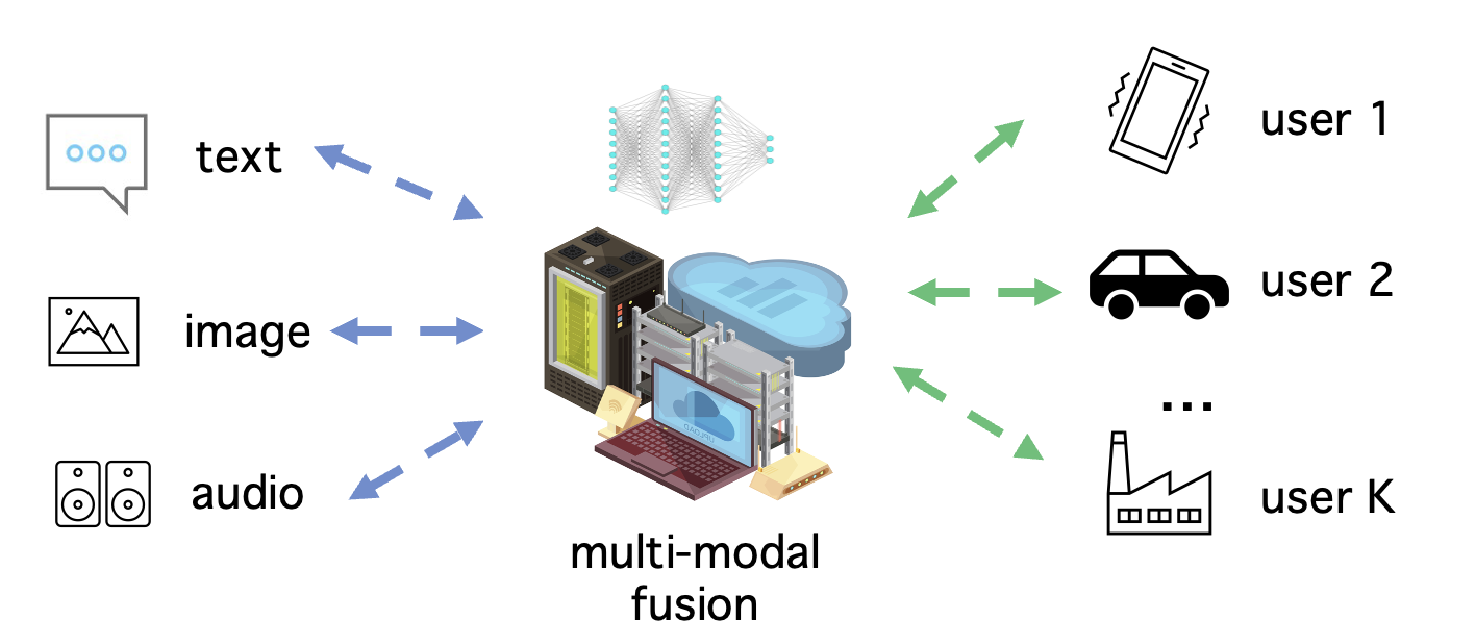}
\caption{Illustration of the multi-modal SemCom system.}
\label{fig:multimodal}
\end{figure}

The evolution of SemCom has progressed from transmitting isolated words or images to integrating multiple modalities, such as text, images, audio, and video. This advancement has given rise to multi-modal SemCom systems, which are designed to handle and process diverse types of data and tasks. As illustrated in Fig.~\ref{fig:multimodal}, a typical multi-modal SemCom system involves several key steps: first, the collection of data from different modalities; next, a central fusion process that integrates these heterogeneous inputs into a unified semantic representation; and finally, the encoded semantic message is distributed to multiple users.
 
One such system is the task-oriented Multi-User SemCom for Visual Question Answering (VQA) framework, known as MU-DeepSC, which integrates multi-modal data from multiple users to enhance VQA performance~\cite{xie2021task}. By jointly optimizing transmitters and receivers, MU-DeepSC effectively captures correlated features across different modalities, demonstrating resilience against channel variations, particularly in low SNR scenarios. Similarly, the Unified multi-task SemCom (U-DeepSC) system offers a flexible framework for handling multiple tasks across various data modalities~\cite{zhang2024unified}. It employs a dynamic feature adaptation mechanism that adjusts transmitted data based on specific tasks and channel conditions, thereby improving transmission efficiency. Additionally, the Multi-modal Fusion-based Multi-task SemCom (MFMSC) framework utilizes BERT to effectively integrate semantic information from different modalities~\cite{zhu2024multi}. This approach not only enhances performance but also reduces communication overhead in complex multi-task environments.  

Despite these advancements, multi-modal SemCom faces several notable challenges. Communication bandwidth overhead remains a significant concern, as transmitting multi-modal data typically demands substantial bandwidth, potentially causing increased communication overhead and latency issues. Additionally, effectively achieving semantic fusion among heterogeneous data is complex due to the diverse information inherent to different modalities. Developing efficient mechanisms to extract and integrate semantic correlations across modalities remains critical. Furthermore, the complexity associated with multi-task and multi-modal frameworks escalates with the increasing number of tasks and modalities. This heightened complexity consequently raises computational costs and limits feasibility, particularly for small-scale or resource-constrained devices.

Moreover, there remain several unresolved challenges that necessitate further exploration. Interpretability is a crucial issue; as these models grow in complexity, understanding and interpreting their decision-making processes become increasingly challenging yet essential. Additionally, achieving performance trade-offs poses a substantial issue, as selecting multi-modal features can significantly affect individual task performance. This requires careful optimization strategies to prevent performance degradation in specific applications.

Addressing these challenges is essential to advancing the efficiency, adaptability, and scalability of multi-modal SemCom systems. The principles of multi-modal SemCom find particularly compelling applications in satellite networks, where harsh channel conditions and stringent resource constraints exacerbate traditional communication inefficiencies. By leveraging semantic extraction and JSCC, satellite systems can overcome bandwidth limitations and high latency, crucially benefiting applications such as Earth observation, real-time monitoring, and other bandwidth-intensive tasks.

\section{Conclusion and Potential Future Directions}
SemCom represents a transformative shift in 6G network design, addressing critical challenges in SE, heterogeneous service integration and resource-constrained environments. This paper has explored the coexistence of SemCom and BitCom through advanced multiple access techniques, demonstrating how NOMA can dynamically allocate resources for optimal performance. The extension to multi-modal frameworks highlights SemCom’s capability to handle diverse data types, while its application in satellite networks underscores its potential to overcome bandwidth limitations and harsh channel conditions through semantic-aware architectures. 

As future works, several key areas demand further research to fully realize SemCom’s potential. Integrating semantic awareness into the physical layer through AI-native air interface designs could unlock new efficiencies in resource utilization in satellite communications. Additionally, energy-efficient implementations will be critical for scaling SemCom to resource-constrained devices, balancing computational overhead with performance gains. As next-generation communication systems evolve, SemCom has the potential to become a cornerstone of intelligent, context-aware networks, reshaping how information is transmitted and processed in the wireless ecosystems of the future.

\bibliographystyle{IEEEtran}
\bibliography{references}
\end{document}